\documentclass[compactaffiliation]{Interspeech}

\interspeechcameraready

\title{Foundation Model Hidden Representations \\for Heart Rate Estimation from Auscultation}

\author[]{Jingping Nie$^{1,2}$, Dung T. Tran$^{1}$, Karan Thakkar$^{1,3}$, Vasudha Kowtha$^{1}$, Jon Huang$^{1}$, \\Carlos Avendano$^{1}$, Erdrin Azemi$^{1}$, Vikramjit Mitra$^{1}$}{}

\affiliation{}{Apple, USA;
$^{2}$The University of North Carolina at Chapel Hill, USA; \\
$^{3}$Johns Hopkins University}{USA}

\email{jingping@unc.edu, dung\_tran@apple.com, kthakka2@jhu.edu, vkowtha@apple.com, jjhuang@apple.com, cavendano@apple.com, eazemi@apple.com, vmitra@apple.com}


\newcommand{\blue}[1]{\textcolor{black}{#1}}

\usepackage{comment}
\usepackage{cite, url}
\usepackage{amsmath,amssymb,amsfonts}
\usepackage{algorithmic}
\usepackage{graphicx}
\usepackage{textcomp}
\usepackage{xcolor}
\usepackage{booktabs,multirow}
\usepackage{adjustbox}
\usepackage{setspace}
\usepackage{tabularx}
\usepackage{subcaption}
\usepackage[margin=1in]{geometry}

\begin{document}

\maketitle

\begin{abstract}
  Auscultation, particularly heart sound, is a non-invasive technique that provides essential vital sign information. Recently, self-supervised acoustic representation foundation models (FMs) have been proposed to offer insights into acoustics-based vital signs. However, there has been little exploration of the extent to which auscultation is encoded in these pre-trained FM representations. In this work, using a publicly available phonocardiogram (PCG) dataset and a heart rate (HR) estimation model, we conduct a layer-wise investigation of six acoustic representation FMs: HuBERT, wav2vec2, wavLM, Whisper, Contrastive Language-Audio Pretraining (CLAP), and an in-house CLAP model. Additionally, we implement the baseline method from \cite{nie2024model} (which relies on acoustic features) and show that overall, representation vectors from pre-trained foundation models (FMs) offer comparable performance to the baseline. Notably, HR estimation using the representations from the audio encoder of the in-house CLAP model outperforms the results obtained from the baseline, achieving a lower mean absolute error (MAE) across various train/validation/test splits despite the domain mismatch.
\end{abstract}

\section{Introduction}
\label{sec:introduction}
Health-related acoustic sounds have significant potential for health, fitness, and wellbeing. Heart sound auscultation (from phonocardiograms, a.k.a PCGs) is a non-invasive method that offers crucial vital sign data (such as heart rate (HR) information), and machine learning can be used to explore features and enable high-value use cases. Heart rate estimation is crucial as it indicates cardiovascular health and overall well-being, aiding in diagnosing and managing conditions such as heart disease, stress, and fatigue~\cite{lamberts2010heart, fauquet2016heart, fox2007resting}. 
Frequency-domain acoustic features, such as mel-spectrograms and mel-frequency cepstral coefficients (MFCCs), have been widely used in heart sound analysis~\cite{nie2024model, noman2019short, qiu2024study, dia2019heart, li2022heart, hou2023arsteth}. These frequency-domain features have been applied in various tasks, such as heart sound segmentation with deep convolutional neural networks (CNNs)~\cite{noman2019short}, HR estimation using non-negative matrix factorization~\cite{dia2019heart}, and heart sound classification~\cite{li2022heart, qiu2024study, hou2023arsteth}. 
An end-to-end 2D CNN using mel-spectrograms, MFCCs, power spectral density, and root mean square energy was recently proposed to estimate HR with superior performance~\cite{nie2024model}. \blue{In addition, the INTERSPEECH COMputational PARalinguistics challengE (ComParE) audio feature set, a 6373-dimensional representation of an
audio instance extracted using the openSMILE toolkit, is used by researchers for heart sound classification and heart activity detection~\cite{schuller2018interspeech, eyben2010opensmile, humayun2018ensemble, ren2018learning, elbanna2024predicting}.}

Research has shown that using representations from pre-trained acoustic FMs can improve emotion recognition performance compared to the performance when using acoustic features~\cite{mitra2024investigating, mitra2023pre, morais2022speech, nie2025multi}. It has been shown that the layer-wise progression of representations follows an acoustic-linguistic hierarchy, with shallow layers encoding acoustic features, followed by phonetic, word identity, and meaning information in pre-trained wav2vec2~\cite{pasad2021layer}. More recently, self-supervised speech FMs have been explored for insights into acoustics-based cardiorespiratory vital signs and disease detection~\cite{mitra2024pre, panah2023exploring, mathew2024foundation}, but no study has explored how much cardiorespiratory information is captured by such models.

\begin{figure}[t!]
    \centering
    \includegraphics[width=0.9\linewidth]{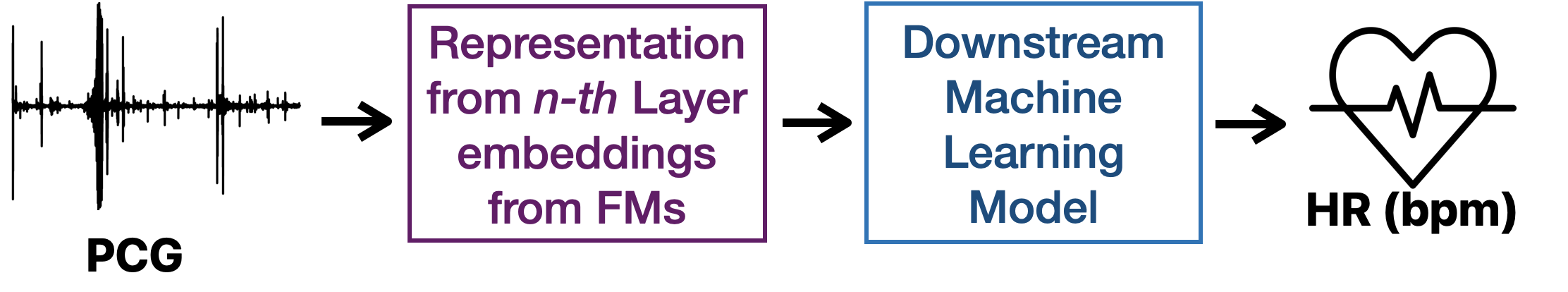}
    \vspace{-3mm}
    \caption{The overall goal of this project.}
    \label{fig:overallgoal}
    \vspace{-0.2 in}
\end{figure}

In this work, as shown in Figure~\ref{fig:overallgoal}, we use a publicly available phonocardiogram (PCG) dataset~\cite{oliveira2022circor}, described in Section~\ref{sec:data}, for a heart rate estimation task. We conduct a layer-wise analysis of the representation vectors extracted from the audio encoders of six self-supervised speech representation FMs, \blue{including five widely recognized speech or audio events FMs}, HuBERT, wav2vec2, wavLM, Whisper, and Contrastive Language-Audio Pretraining (CLAP), as well as an in-house CLAP model.  These models, originally developed for automatic speech recognition (ASR) or audio-language multimodal tasks, present a domain mismatch when applied to vital sign detection. In particular, HuBERT learns spoken language representations via self-supervised learning~\cite{hsu2021hubert}, while wav2vec2 captures contextual speech representations for ASR~\cite{baevski2020wav2vec}. WavLM improves tasks like speaker diarization and speech enhancement~\cite{chen2022wavlm}, and Whisper is robust to noise and varied audio quality~\cite{radford2022whisper}. CLAP (Contrastive Language-Audio Pretraining) is trained on audio-text pairs to predict the most relevant text snippet for a given audio~\cite{elizalde2023clap}. Like CLAP, our in-house CLAP uses contrastive training to align the embedding space between the audio encoder and a text encoder. Our audio encoder is pre-trained using self-supervision following a similar idea as AudioMAE~\cite{huang2022masked}. The details of the in-house CLAP model are described in Section~\ref{sec:model_clap}. 

Additionally, we employ the acoustic features-based method proposed in~\cite{nie2024model} as the baseline to assess whether representations from pre-trained FMs, despite domain mismatch, offer superior performance. We observed that using the representation vectors from pre-trained FMs achieves a comparable performance compared to the baseline method using acoustic features, while sometimes having a more stable performance with different train/validation/data splits. Using the representations from the in-house CLAP model outperforms using the acoustic features in the baseline approach, achieving a lower mean absolute error ($MAE$) across various train/validation/test splits despite the domain mismatch. \blue{This work establishes a foundational benchmark for future research and potential adaptations of FMs in heart sound analysis, broader cardiorespiratory sound assessment, and clinical applications. Given the widespread adoption of these FMs across diverse domains, our findings can be seamlessly integrated into existing FM-powered systems to enhance cardiovascular monitoring and provide valuable health insights.}

\section{Data}
\label{sec:data}
\vspace{-0.05in}
The PCG dataset used in our study is the \emph{CirCor DigiScope Phonocardiogram dataset}, containing 3,163 heart sound recordings from 942 subjects \blue{collected across four main auscultation sites in hospitals. Each PCG recording spans from 5.1 to 64.5 seconds, totaling about 20 hours.}
All recordings in this dataset are low-pass filtered with a cutoff frequency of $2,000\thinspace Hz$. The segmentation annotations (onsets and offsets of S1 and S2 sounds) in this dataset were derived using a semi-supervised method that employed a voting mechanism among three machine-learning approaches and verified by an expert annotator and \blue{the cardiac murmurs were annotated by a human annotator}~\cite{oliveira2022circor}. \blue{These annotations were independently generated and were not influenced by our work and recognition of heart murmur is not the goal of this work.} This \emph{CirCor} PCG dataset presents challenges for robust HR estimation due to biases and errors in segmentation annotations, as well as unannotated noises (e.g., environmental background noises) \blue{and unannotated HR-related pathological conditions (e.g., arrhythmia)}. 

\begin{figure}[t!]
    \centering
    \includegraphics[width=0.45\textwidth]{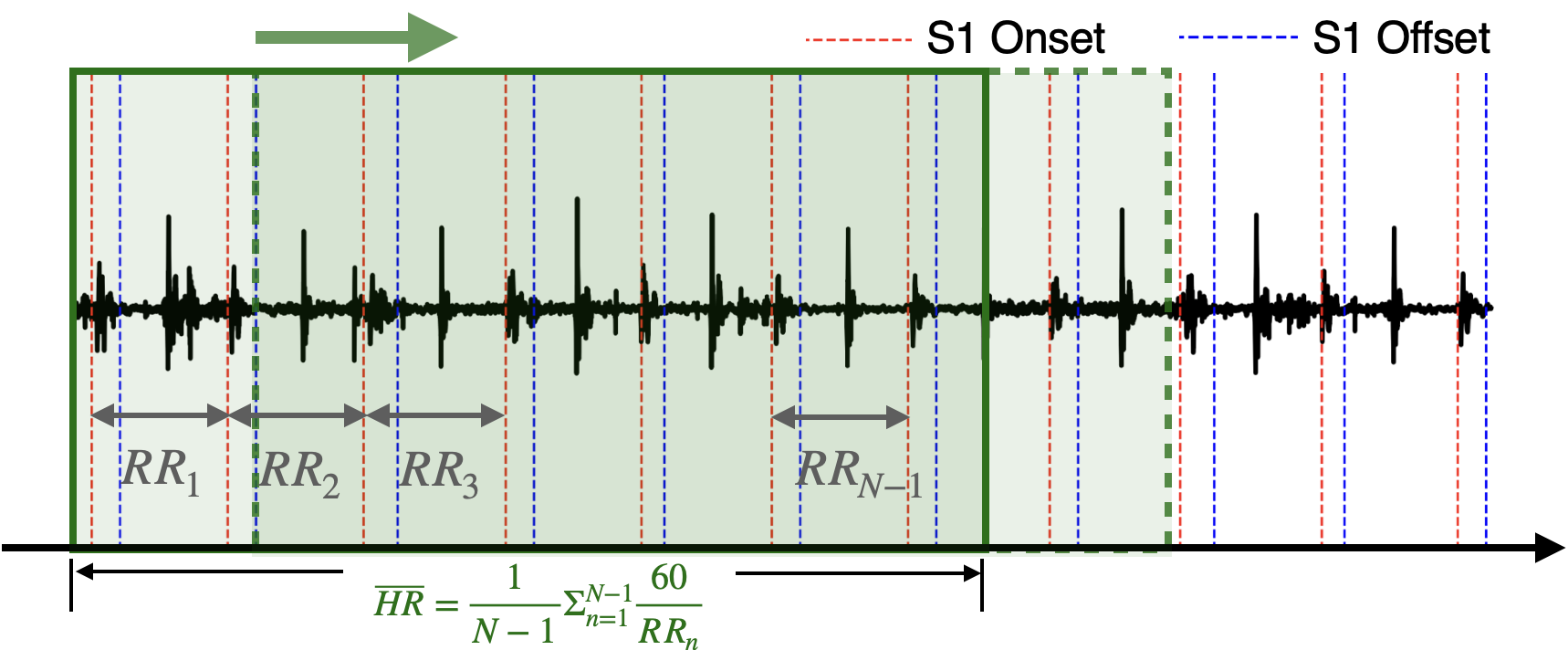}
    \vspace{-0.5\baselineskip}
    \caption{The data preparation process.}
    \vspace{-\baselineskip}
    \label{fig:training_preparation}
    
\end{figure}

\begin{figure}[t!]
    \centering
    \includegraphics[width=0.49\textwidth]{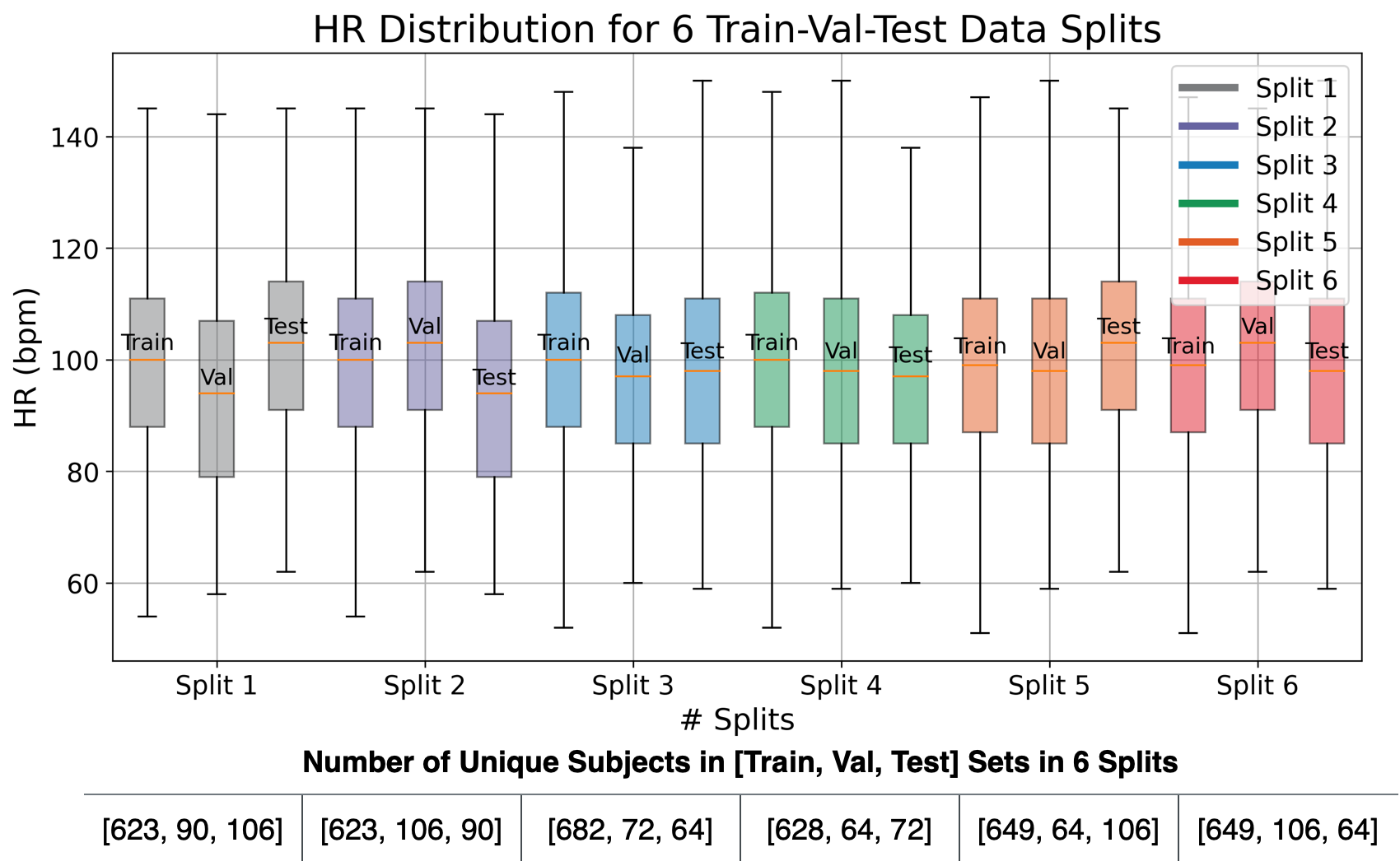}
    \vspace{-0.2in}
    \caption{The heart rate distribution for 6 training, validation, and test splits and the number of unique subjects.}
    \vspace{-0.15in}
    \label{fig:splits}
\end{figure}

Similar to \cite{nie2024model}, a sliding window of $5\thinspace s$ with a stride of $1\thinspace s$ is applied to raw PCG audio files where the annotated period exceeds $5\thinspace s$ (see Figure~\ref{fig:training_preparation}) \blue{to ensure a fair comparison with the acoustic feature-based method proposed in~\cite{nie2024model} and to generate a sufficient number of sound snippets for model training in this study}. This process generates $23,381$ heart sound snippets. Average heart rate for each sound snippet is generated using the method proposed in \cite{nie2024model}, where the average heart rate ($\overline{HR}$) in beats per minute (bpm) is based on the interval between adjacent S1 onsets as illustrated in Figure~\ref{fig:training_preparation}. The dataset is then split into training ($80\%$), validation ($10\%$), and test ($10\%$) sets using 6 different splits. This approach helps assess model robustness, mitigate overfitting, and comprehensively evaluate the FMs investigated in this study. Importantly, no overlap exists between the training, validation, and test sets, and no subject appears in more than one of these splits. The heart rate distribution for training, validation, and test sets in 6 splits and the number of unique subjects in each set are illustrated in Figure~\ref{fig:splits}.  

\begin{figure*} [t!]
    \centering
    \includegraphics[width=0.9\textwidth]{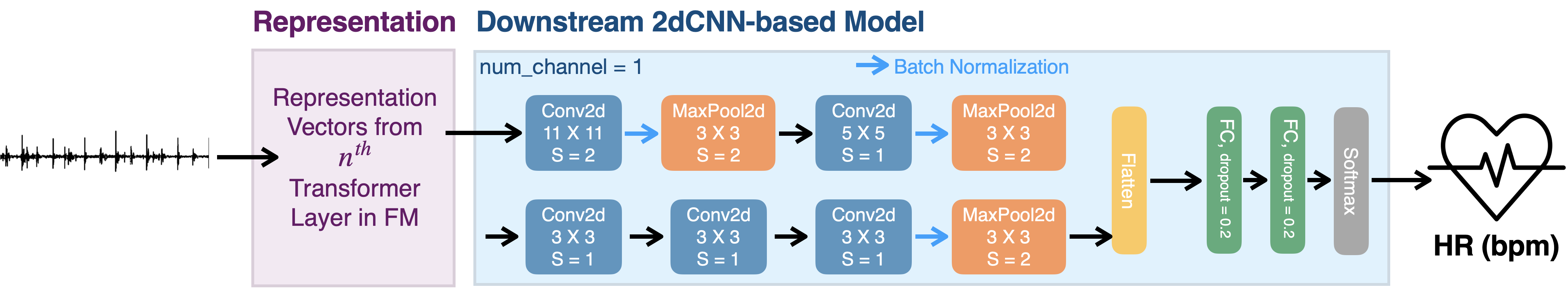}
    \caption{The representation vectors of $n^{th}$ embedding layer in the audio encoder of the foundation model are passed into a downstream 2D convolutional neural network (\textbf{\texttt{2dCNN}}) for HR estimation.}
    \label{fig:AlexNetHR}
\end{figure*}

\begin{table*}[h]
\centering
\caption{Performance comparison across the FMs and baseline approach.}
\vspace{-0.1in}
\resizebox{0.99\textwidth}{!}{
\begin{tabular}{l|c|c|c|c|c|c|c}
\hline
\textbf{Model} & Dimension & \#Parameters &$\min(\overline{MAE})$& \textbf{$\sigma_{MAE_{j*}}$} & $\underset{i,j}{\min}\left(MAE_{i,j}\right)$ & \textbf{$\frac{1}{N} \sum_{j=0}^{N} \sigma_{MAE}$} & $\min\left(\sigma\right)$ \\ \hline\hline
HuBERT Large & Layer 1-24: [249,1024] & $\sim$ 316M & 2.23  & 0.44 & 1.53  & 0.55 & 0.35 \\ \hline
HuBERT Base & Layer 1-12: [249,768] &$\sim$ 95M & 2.26  & 0.26 & 1.77  & 0.56 & 0.26 \\ \hline
Wav2Vec2 Large & Layer 1-24: [249,1024] & $\sim$ 317M &2.49  & 0.64 & 1.86  & 0.73 & 0.36 \\ \hline
Wav2Vec2 Base & Layer 1-12: [249,768] & $\sim$ 95M&2.41  & 0.48 & 1.79  & 0.80 & 0.40 \\ \hline

WavLM Large & Layer 1-24: [249,1024] & $\sim$ 315M &2.02  & 0.25 & 1.85  & 0.41 & 0.41 \\ \hline
WavLM Base & Layer 1-12: [249,768] & $\sim$ 94M &2.20  & 0.51 & 1.59  & 0.52 & 0.34 \\ \hline
Whisper Large & Layer 1-32: [250,1280] & $\sim$ 1.55B &2.27  & 0.37 & 1.62  & 0.78 & 0.27 \\ \hline
Whisper Small & Layer 1-12: [250,1280] & $\sim$ 39M &2.27  & 0.51 & 1.67  & 0.81 & 0.33 \\ \hline
in-house CLAP & Layer 1-12: [248,768] & $\sim$ 85M &\textbf{1.88 } & 0.37 & 1.33  & 0.49 & 0.37 \\ \hline
MSCLAP & \begin{tabular}[c]{@{}c@{}}Layer 1: [1024,192] \\ Layer 2: [256,384] \\ Layers 3 and 4: [64,768]\end{tabular} & $\sim$ 159M &2.84 & 0.61 & 1.89 & 0.61 & 0.55 \\ \hline\hline
Wav2Vec2 Base ASR & Layer 1-12: [249,768] & $\sim$ 95M &2.40  & 0.37 & 1.95  & 0.91 & 0.36 \\ \hline
\hline
Baseline & \multicolumn{7}{c}{Average $MAE$: 1.91; $STD$: 0.32} \\ \hline
\end{tabular}
}
\vspace{-0.2in}
\label{tab:performance_comparison}
\end{table*}

\section{In-House CLAP Model}\label{sec:model_clap}

\textbf{Audio Encoder Architecture:} The audio encoder takes a 128 LogMel spectrogram as input, computed from mono 16 kHz raw audio using a 25 ms Hann window and 10 ms hops. The model is a 12-layer ViT-B with around 86 million parameters. A 16×16 non-overlapping Conv2D layer is used for patch embedding in the first layer of the ViT before flattening it into a 1D sequence. For a 10-second input audio, the spectrogram has dimensions of 1×128×1024, and after patch embedding, it results in a 1×512 sequence. This sequence of tokens is then fed into a 12-layer transformer with 768 hidden dimensions and additive sinusoidal positional encoding. The output of the final transformer layer is mean-pooled to generate a single 768-dimensional embedding for the entire input audio.

\noindent\textbf{Audio Encoder Pre-training:} The audio encoder is pre-trained using a self-supervised approach, similar to AudioMAE~\cite{huang2022masked}, with an encoder-decoder architecture. We use a 10-layer Conv1D decoder with 768-dimensional representations and a kernel size of 7. The decoder’s output is depatchified to produce the original 1×128×1024 spectrogram. During training, 80\% of the patch embeddings are randomly masked at the input of the audio encoder transformer, and the combined network is tasked with reconstructing the unmasked spectrogram. We train our model on an internal dataset of approximately 3 million audio samples for 100 epochs with an effective batch size of 2048. A one-cycle learning rate schedule with a peak of 1e-4 and the AdamW optimizer is used. Since this training step focuses on learning robust representations, the encoder is retained for downstream tasks, while the decoder is discarded.

\noindent\textbf{Audio-text Contrastive Training:} We use contrastive training to align the embedding space between the previously described audio encoder and a text encoder. Our dataset of approximately 3 million audio samples contains around 5 million human-generated captions. The outputs of both the audio and text encoders pass through a linear projection layer to 512 dimensions. We trained both encoders for 30,000 iterations with an effective batch size of 8,192.
\vspace{-0.05in}

\section{Downstream Model and Training}\label{sec:model}
Since HR is typically represented as an integer in beats per minute, we followed a similar approach to~\cite{nie2024model}, treating HR estimation as a 141-class classification problem, where $\overline{HR} \in[40, 180]$ (bpm). The cross-entropy (CE) loss function was used for HR estimation:
\begin{eqnarray}
    &CE = \sum_{a=1}^{A}(-\sum_{b=1}^{B}\log\frac{\exp{(x_{a,c})}}{\sum_{i = 1}^{B}\exp{(x_{n,i})}}y_{a,b}), 
\end{eqnarray}
where $x$ is the input, $y$ is the target, $A$ is the number of audio snippets, $B$ is the number of HR estimation classes. Mean absolute error ($MAE$) was used to evaluate the model performance for the HR estimation: 
\begin{eqnarray}
    MAE = \frac{1}{M}\sum_{m=1}^{M}|\overline{HR}_{predicted, m} - \overline{HR}_{target, m}|, 
\end{eqnarray}
where $M$ is the total number of samples in the dataset.

As shown in Figure~\ref{fig:AlexNetHR}, the representation vectors from the $n^{th}$ embedding layer of the audio encoders in the FM are extracted from 5-second long sound snippets and passed into a downstream 2D convolutional neural network (CNN) classification model inspired by~\cite{nie2024model} to estimate HR. The downstream model consists of five convolutional layers and three max-pooling layers with varying filter sizes and strides. Batch normalization is applied before each max-pooling layer, and the rectified linear unit (ReLU) is used as the activation function. The multi-dimensional output from the convolutional and pooling layers is flattened into a one-dimensional vector, which is then passed to the final fully connected layers for HR classification. For the representations from each foundation model embedding layer, the downstream model was trained with a mini-batch size of $32$ for $50$ epochs for all dataset splits using Adam as the optimizer.
\vspace{-0.1in}
\begin{figure}[h!]
    \centering
    \begin{tabular}{cc}
        \begin{subfigure}{0.49\linewidth}
            \centering
            \includegraphics[width=\linewidth]{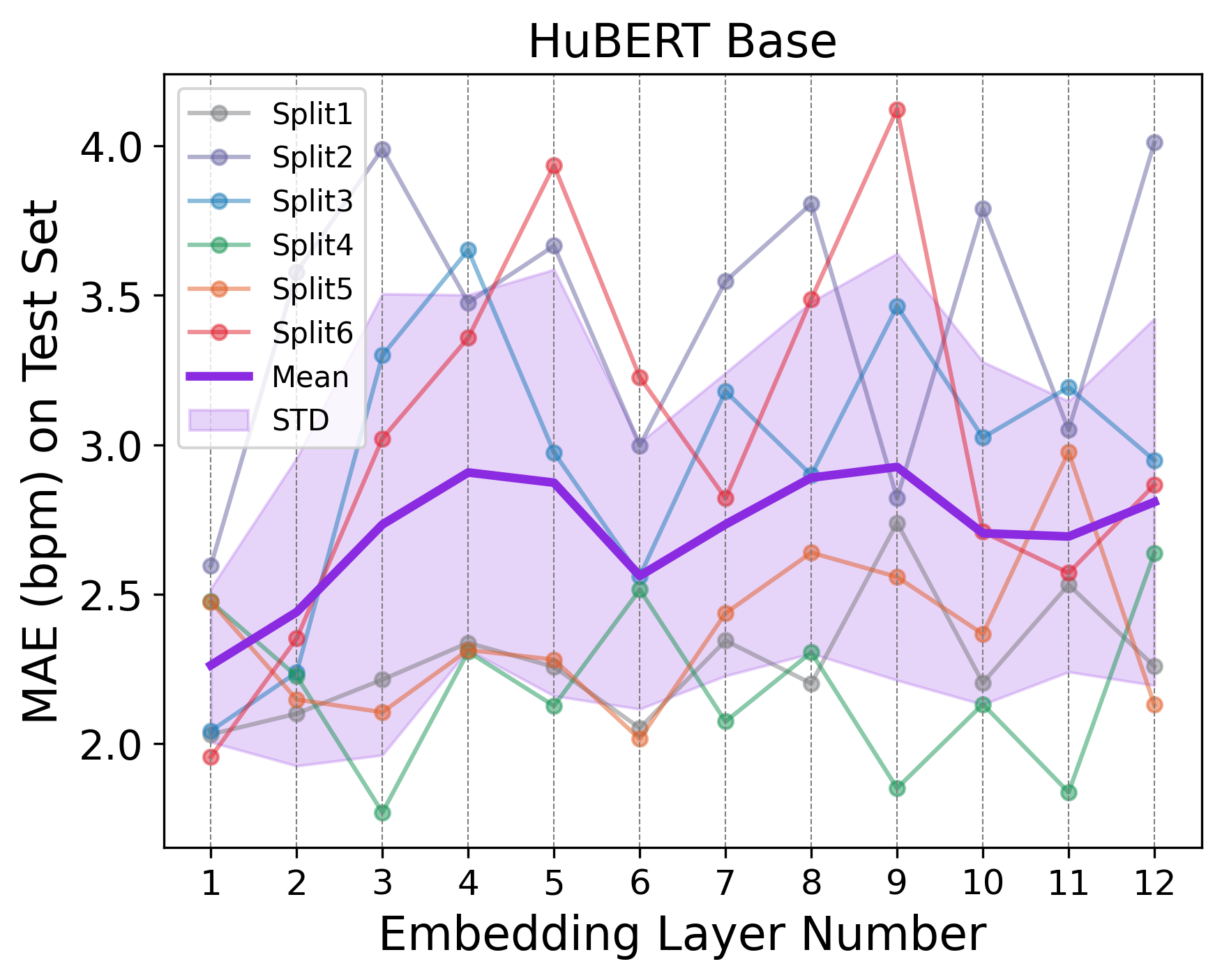}
            \vspace{-0.1in}
            \caption{HuBERT Base}
            \label{fig:HuBERT_Base}
        \end{subfigure} &
        \begin{subfigure}{0.49\linewidth}
            \centering
            \includegraphics[width=\linewidth]{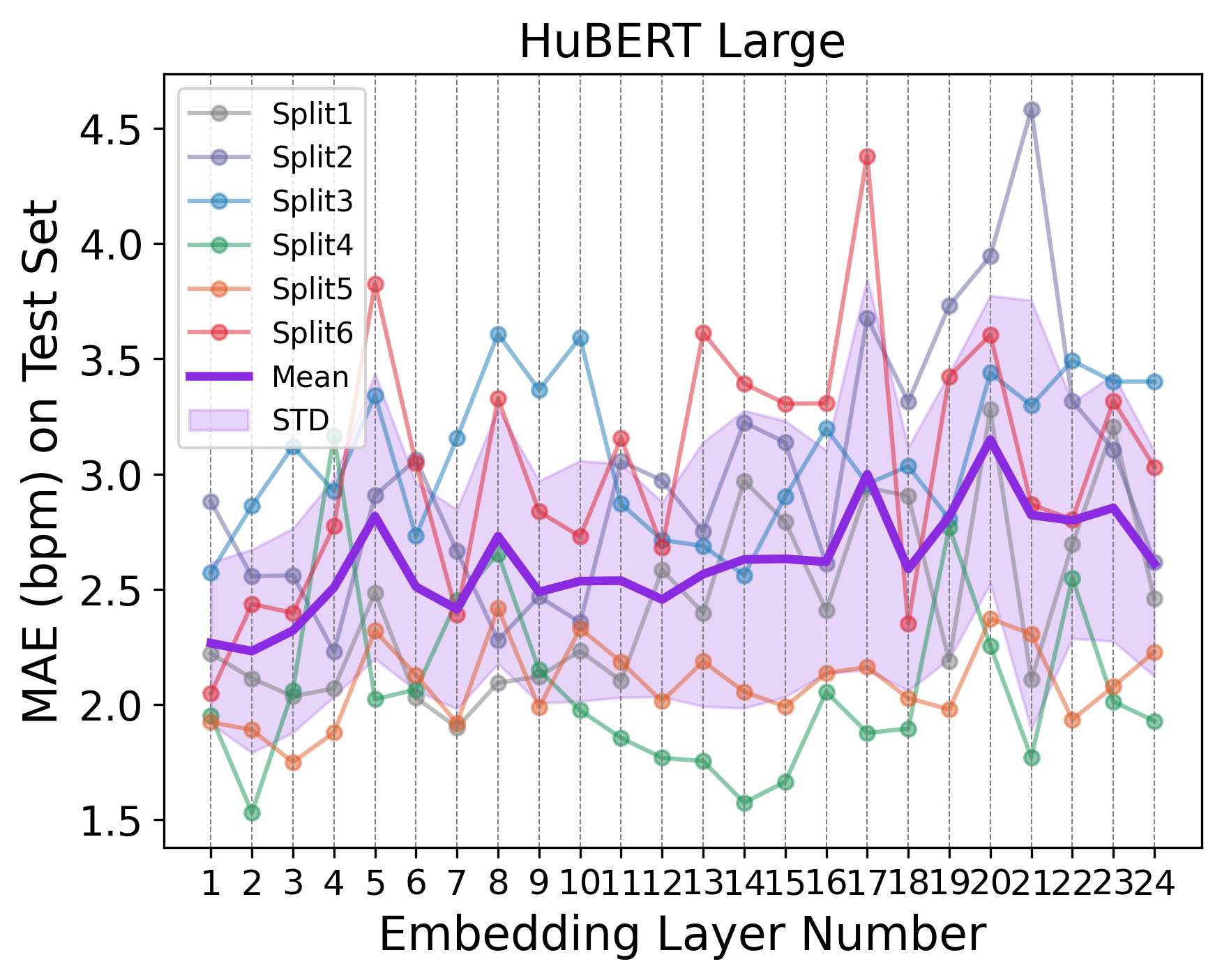}
            \vspace{-0.1in}
            \caption{HuBERT Large}
            \label{fig:HuBERT_Large}
        \end{subfigure} \\
        
        \begin{subfigure}{0.49\linewidth}
            \centering
            \includegraphics[width=\linewidth]{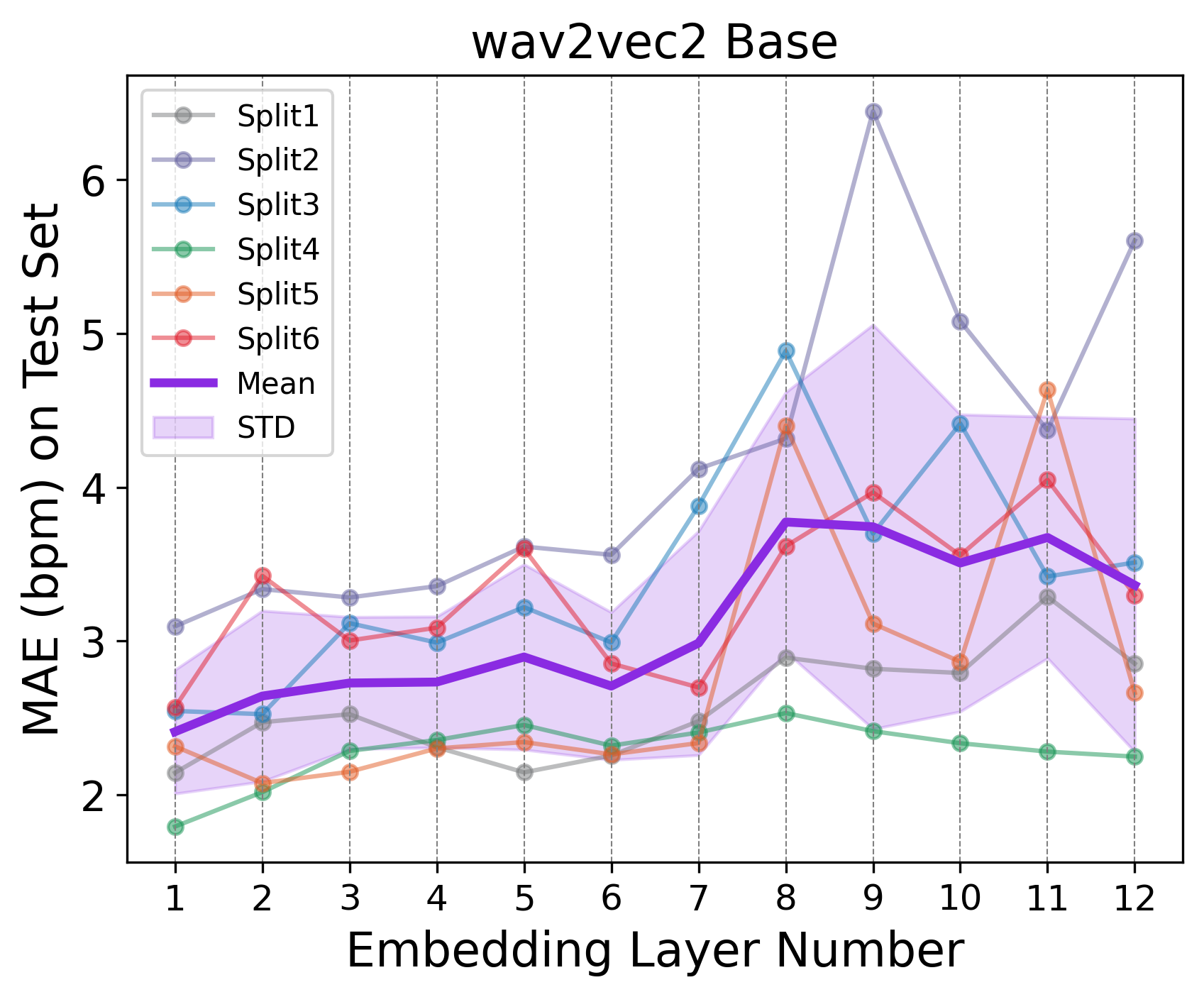}
            \caption{wav2vec2 Base}
            \label{fig:wav2vec2_Base}
        \end{subfigure} &
        \begin{subfigure}{0.49\linewidth}
            \centering
            \includegraphics[width=\linewidth]{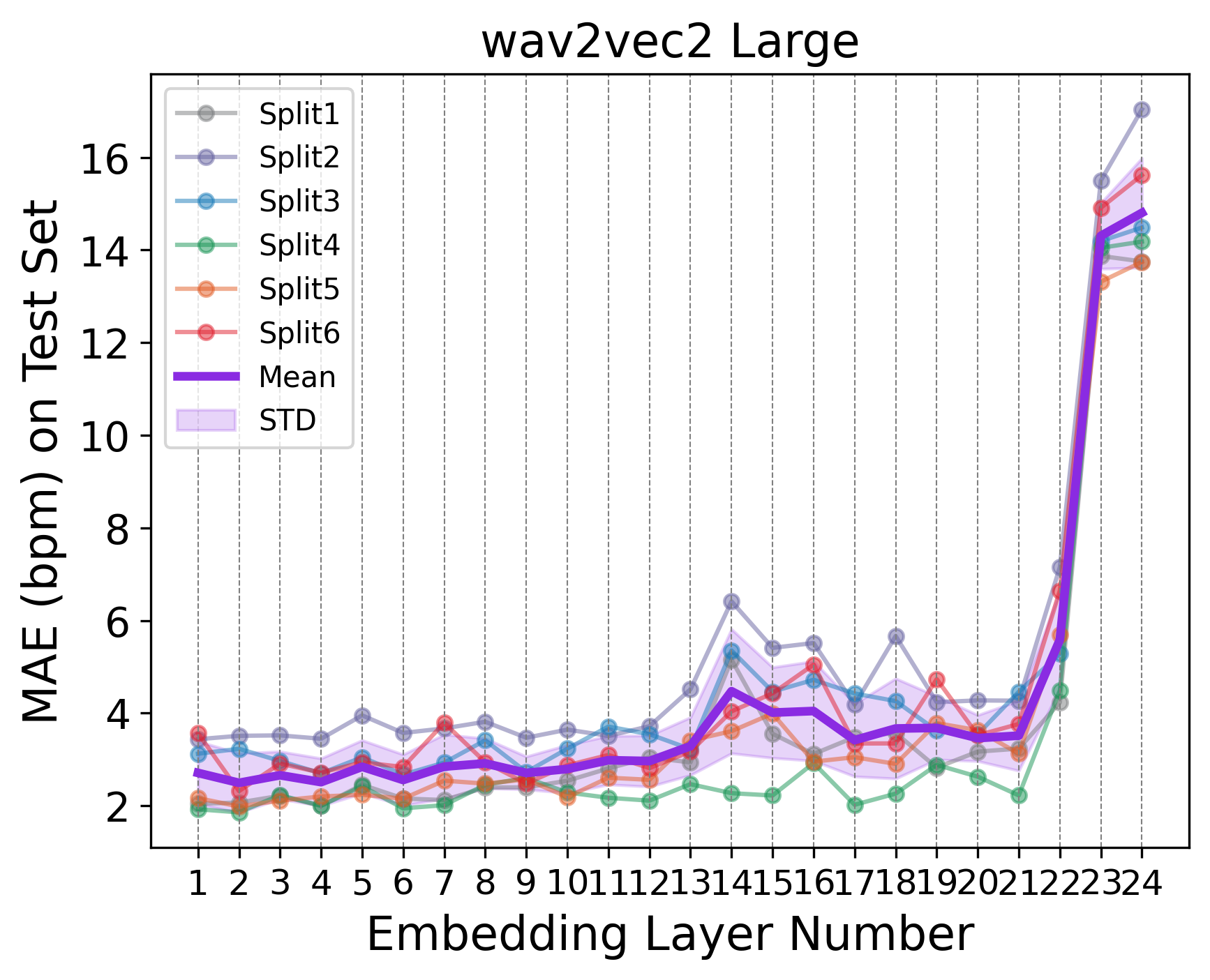}
            \caption{wav2vec2 Large}
            \label{fig:wav2vec2_Large}
        \end{subfigure} \\
        
        \begin{subfigure}{0.49\linewidth}
            \centering
            \includegraphics[width=\linewidth]{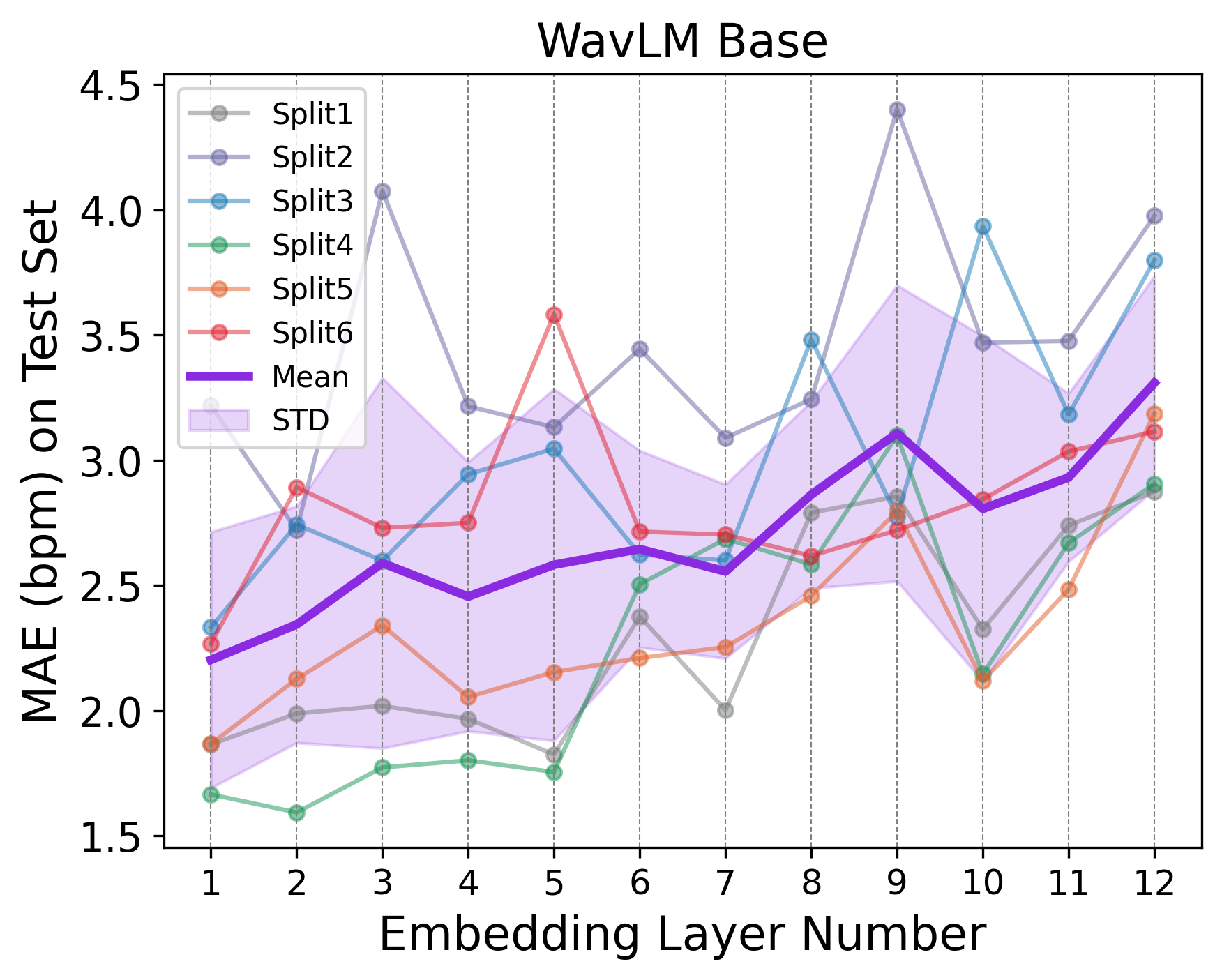}
            \caption{WavLM Base}
            \label{fig:WavLM_Base}
        \end{subfigure} &
        \begin{subfigure}{0.49\linewidth}
            \centering
            \includegraphics[width=\linewidth]{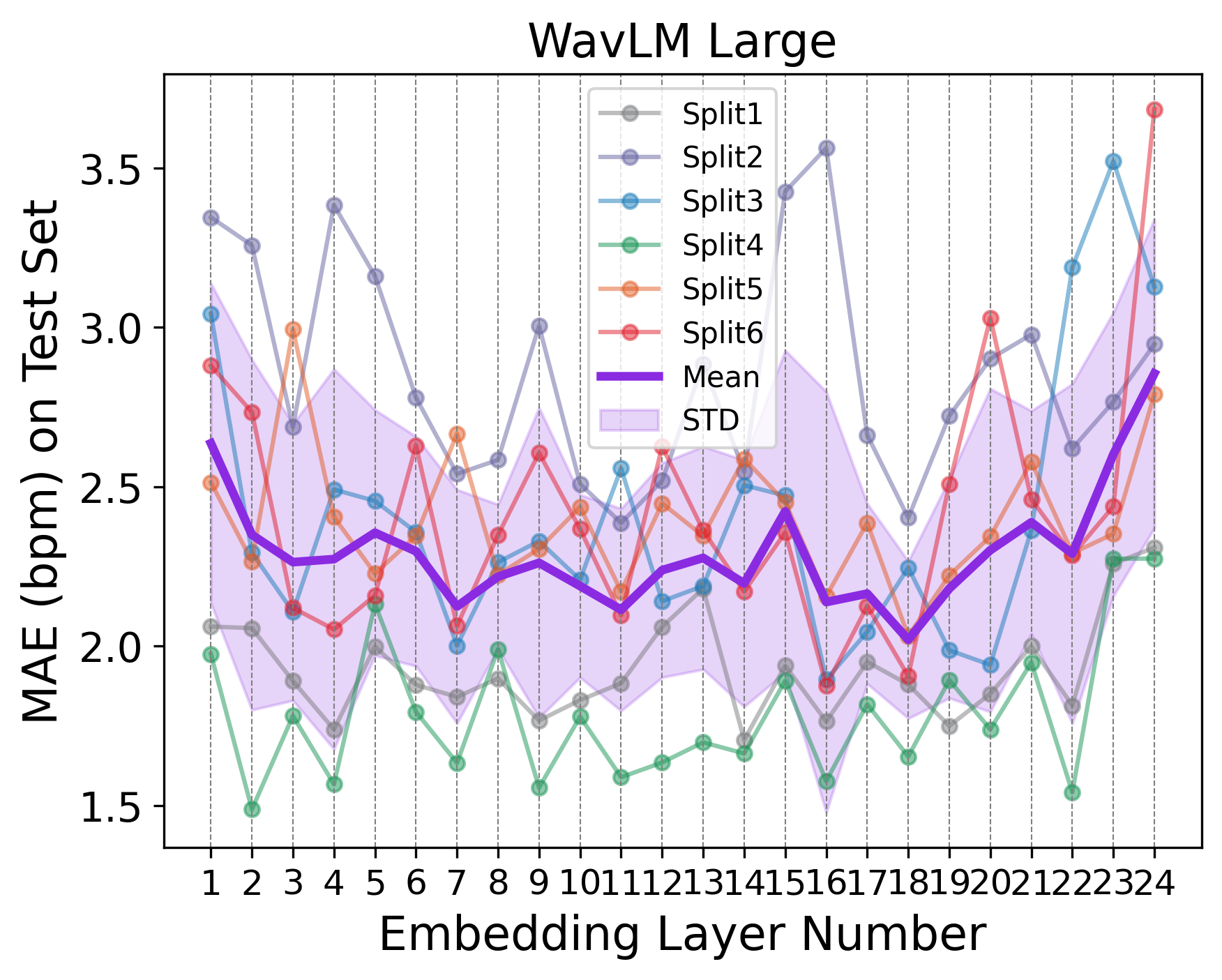}
            \caption{WavLM Large}
            \label{fig:WavLM_Large}
        \end{subfigure} \\
        
        \begin{subfigure}{0.49\linewidth}
            \centering
            \includegraphics[width=\linewidth]{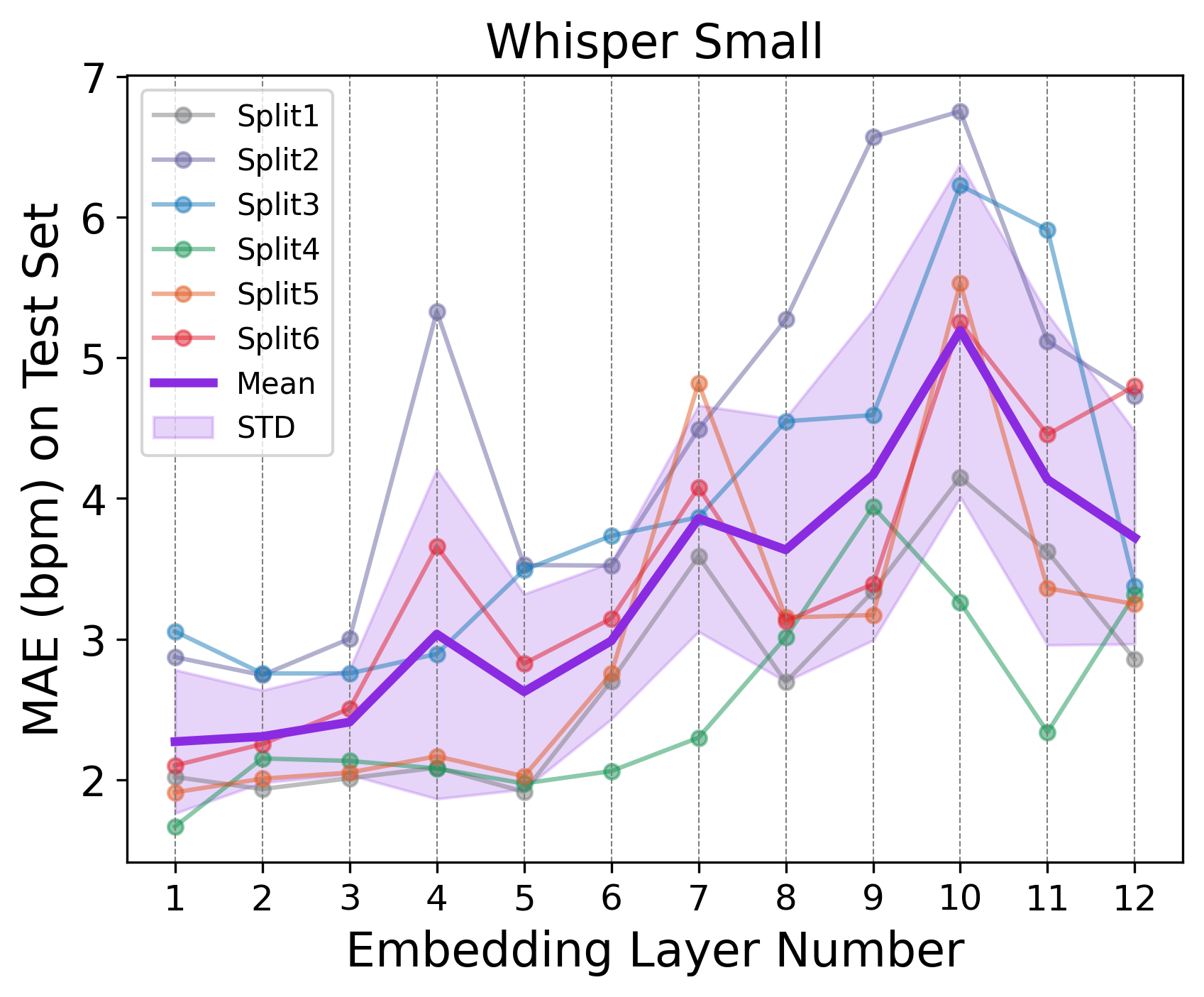}
            \caption{Whisper Small}
            \label{fig:Whisper_Small}
        \end{subfigure} &
        \begin{subfigure}{0.49\linewidth}
            \centering
            \includegraphics[width=\linewidth]{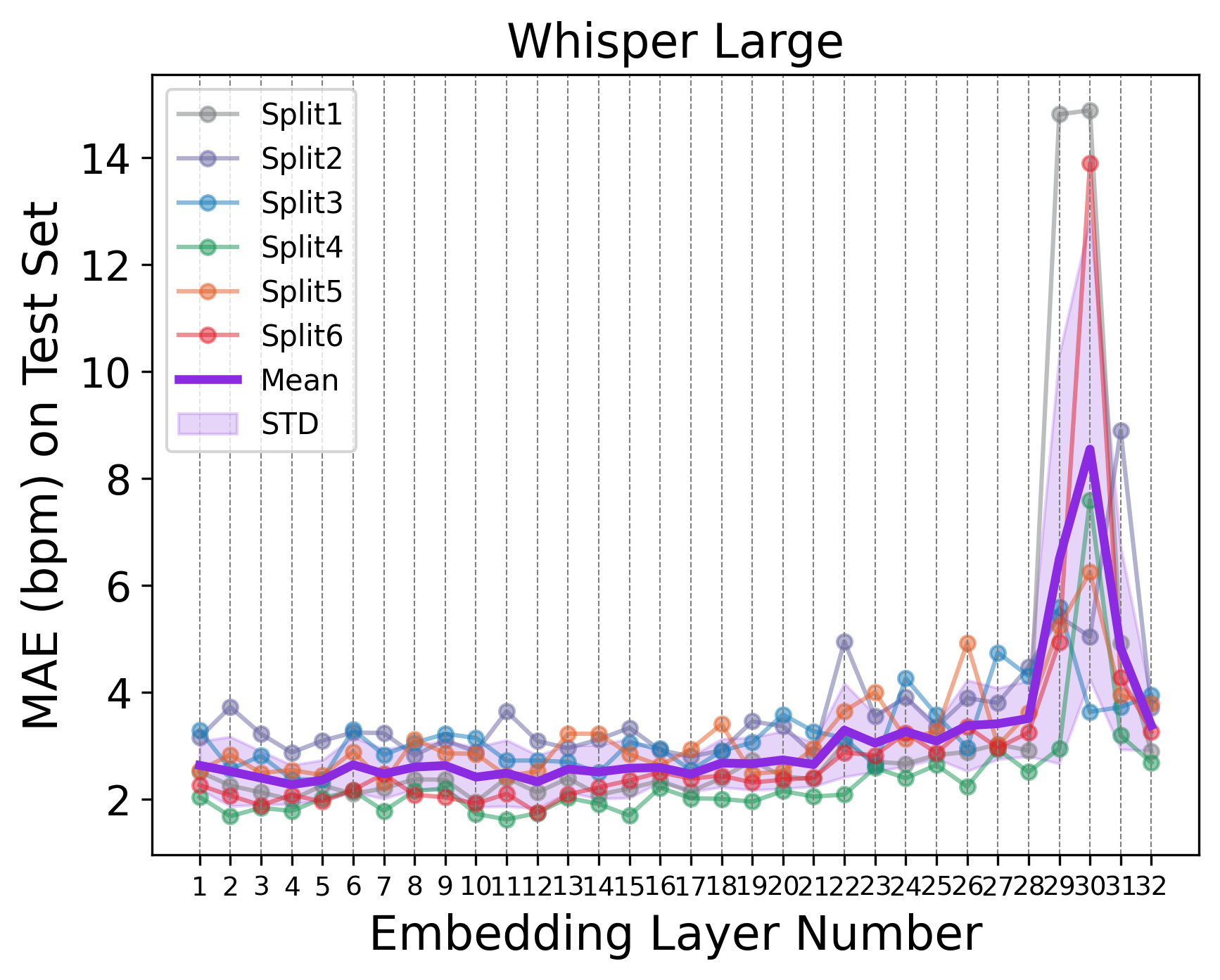}
            \caption{Whisper Large}
            \label{fig:Whisper_Large}
        \end{subfigure} \\
        
        \begin{subfigure}{0.49\linewidth}
            \centering
            \includegraphics[width=\linewidth]{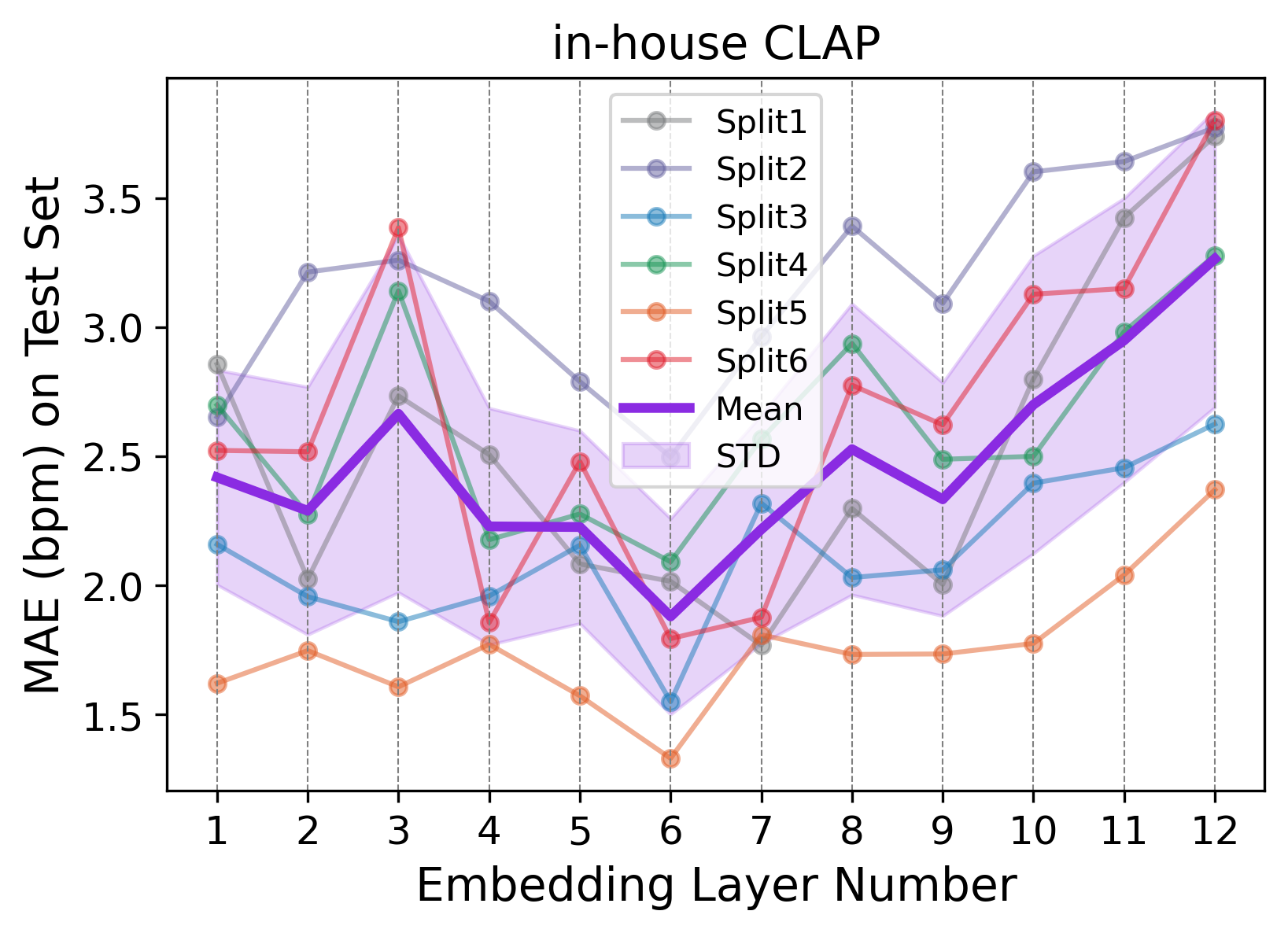}
            \caption{in-house CLAP}
            \label{fig:SoundPrint_CLAP}
        \end{subfigure} &
        \begin{subfigure}{0.45\linewidth}
            \centering
            \includegraphics[width=\linewidth]{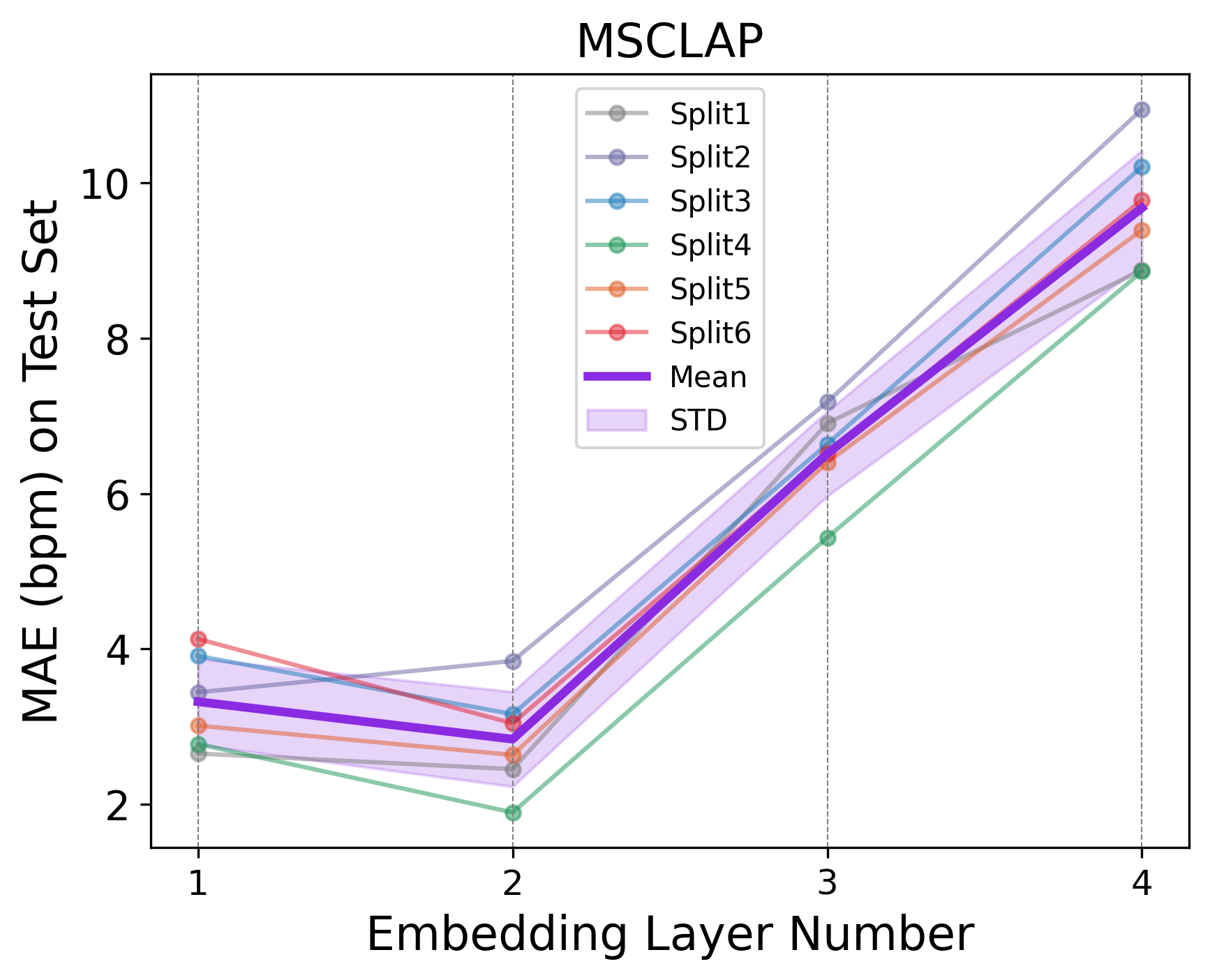}
            \caption{MSCLAP}
            \label{fig:MSCLAP}
        \end{subfigure} \\
    \end{tabular}
    \vspace{-0.1in}
    \caption{The Mean Absolute Error ($MAE_{i,j}$) across six data splits for different feature models (FMs) at various embedding layers.}
    \label{fig:mainfig}
    \vspace{-0.2in}

\end{figure}
\section{Results}\label{sec:result}

Figure~\ref{fig:mainfig} shows the $MAE_{i,j}$ for split $i$ using the representations from embedding layer $j$ of the different FMs. The thick purple line represents the average MAE ($\overline{MAE}j$) across the 6 splits, with the light purple shaded area indicating the standard deviation ($\sigma_{MAE_j}$). 
Table~\ref{tab:performance_comparison} further illustrates the performance of using the representations from different FMs. In particular, (\emph{i}) $\min(\overline{MAE})$ represents the smallest mean across all embedding layers; (\emph{ii}) \textbf{$\sigma_{MAE_{j*}}$} indicates the standard deviation of the MAE values across the different splits at the embedding layer where the mean MAE is the lowest; (\emph{iii}) $\underset{i,j}{\min}\left(MAE_{i,j}\right)$ is the smallest individual MAE value among all the embedding layers and all splits; (\emph{iv}) \textbf{$\frac{1}{N} \sum_{j=0}^{N} \sigma_{MAE}$} corresponds to the average of the standard deviations across all embedding layers; and (\emph{v}) $\min\left(\sigma\right)$ represents the smallest standard deviation among all the embedding layers. Moreover, using the same 6 train/validation/test set splits, the baseline method using acoustic features achieves an average $MAE$ of 1.91 (bpm) with the $STD$ being 0.32.  

As shown in Figure~\ref{fig:mainfig} and Table~\ref{tab:performance_comparison}, the representation vectors from pre-trained FMs generally achieve performance comparable to the baseline method using acoustic features. Notably, the $6^{th}$ embedding layer of the in-house CLAP model achieves an average $MAE$ of 1.88 bpm, outperforming the baseline. The in-house CLAP model demonstrated superior performance over the baseline, which can be attributed to its distinct training objective. Unlike HuBERT, wav2vec2, wavLM, or Whisper, which were predominantly trained for automatic speech recognition tasks, the in-house CLAP model was trained for audio event detection encompassing a wider variety of audio classes beyond speech (as discussed in Section~\ref{sec:model_clap}). This broader training focus likely enhances its ability to capture non-speech features relevant to heart sounds, contributing to its improved effectiveness. The improved performance of the $6^{th}$ layer further suggests that mid-level representations in CLAP strike a balance between low-level acoustic cues and high-level contextual information.

To address practical deployment concerns, we further include model size (i.e., number of parameters) in Table~\ref{tab:performance_comparison}. Although using FM representations with a downstream classifier introduces more parameters compared to the baseline method using acoustic features as the representations, this configuration provides notable advantages. FM embeddings, pretrained on diverse audio data, generalize better across individuals and are reusable across tasks without retraining. In particular, some FMs, like HuBERT Base and WavLM Large, show smaller variances across different train/validation/test splits compared to the baseline. This observation suggests that these models may offer more consistent representations across diverse subjects and varying data distributions. 

\begin{figure}[t!]
    \begin{minipage}{0.45\linewidth}
        \centering
        \includegraphics[width=\linewidth]{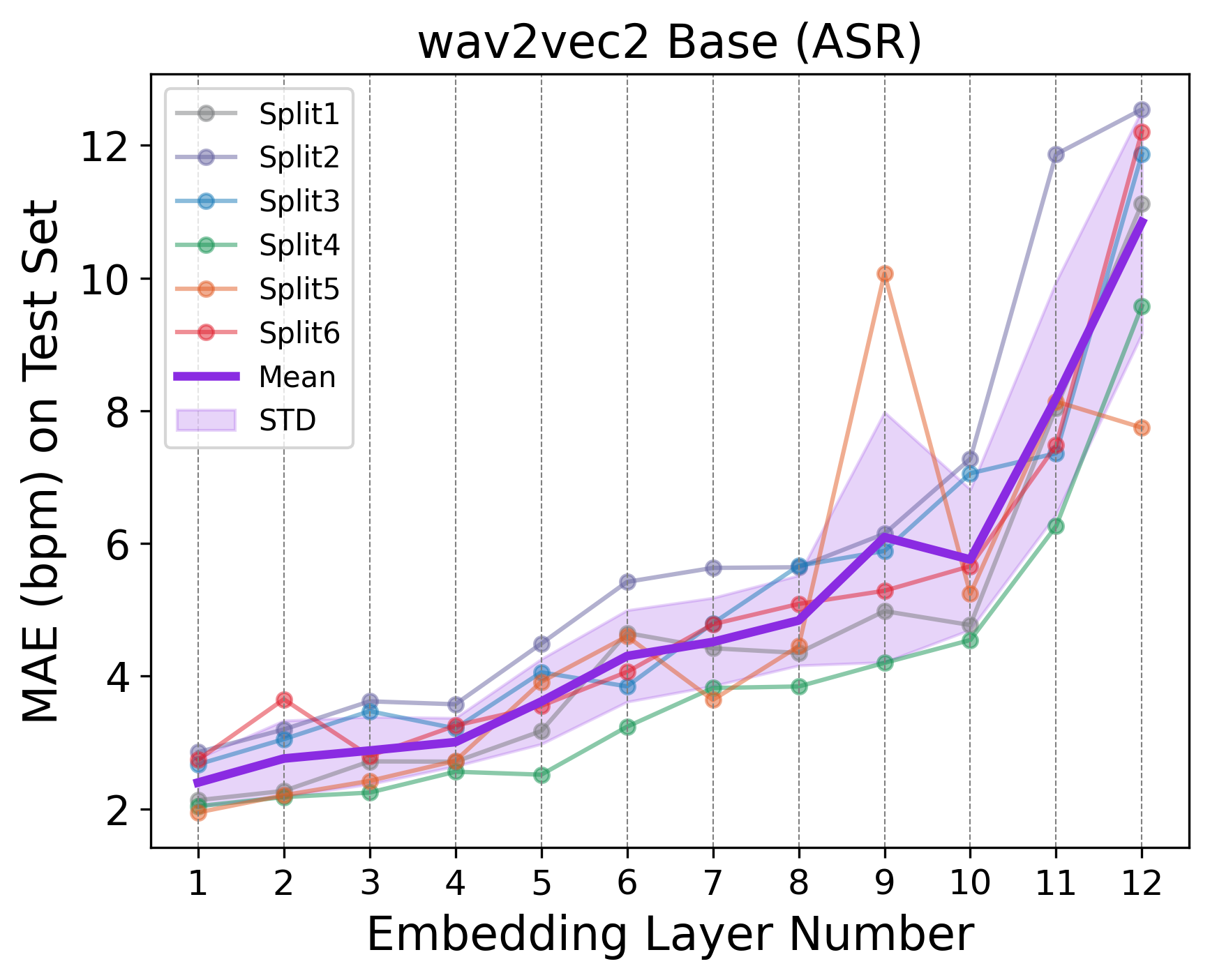}
        \vspace{-0.1in}
        \caption{wav2vec2 Base fine-tuned on 960 hours of Librispeech.}
        \label{fig:wav2vec2_ASR}
    \end{minipage}
    \hfill
    \begin{minipage}{0.5\linewidth}
        \centering
        \includegraphics[width=\linewidth]{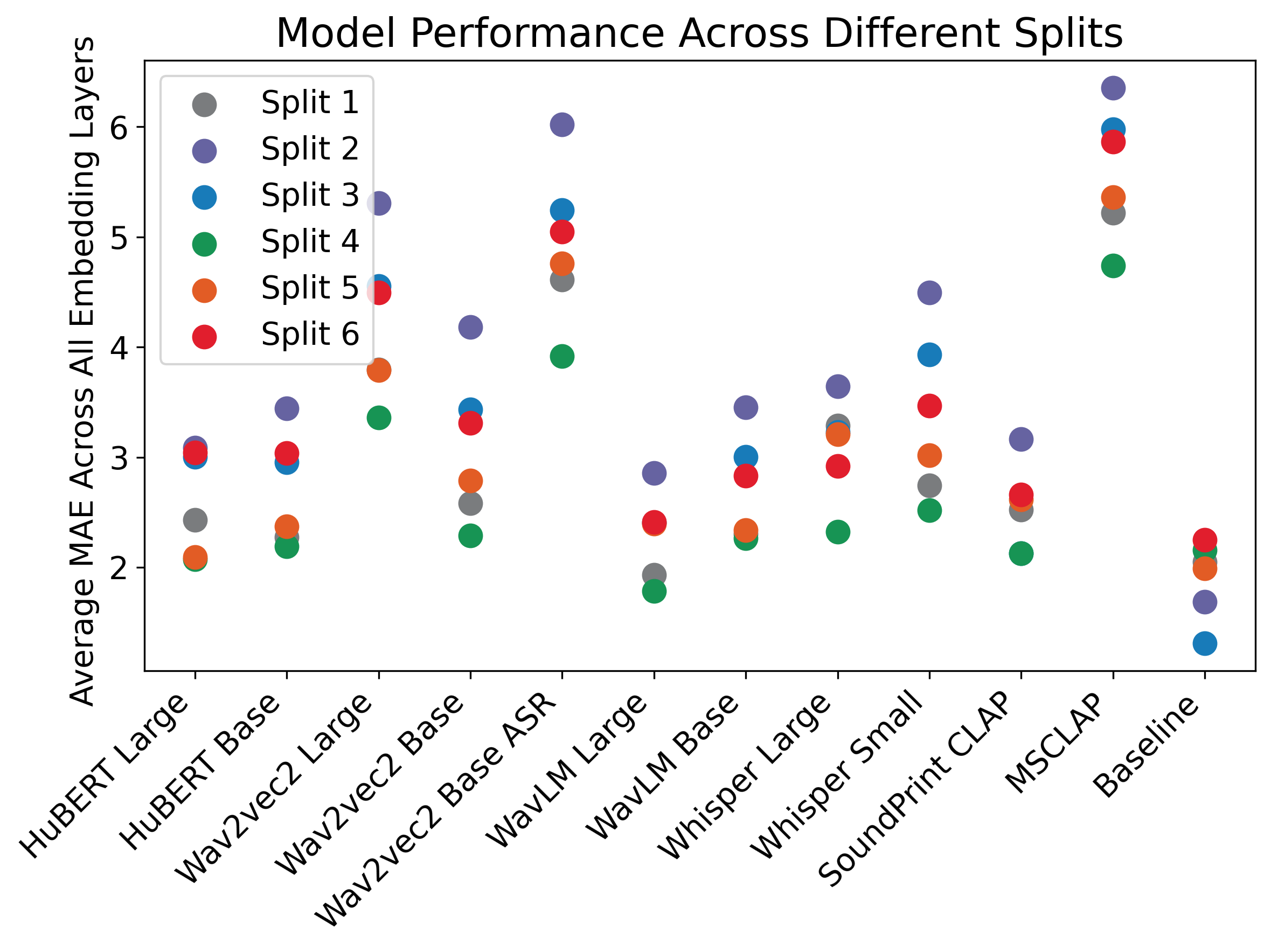}
        \vspace{-0.1in}
        \caption{Overall model performance across 6 train/validation/test splits.}
        \label{fig:ModelPerformance_Splits}
    \end{minipage}
\vspace{-0.2in}
\end{figure}

We found that shallower layers of ASR-based FMs generally perform better in HR estimation. As illustrated in Figure~\ref{fig:wav2vec2_ASR}, the deeper layers of the Librispeech fine-tuned wav2vec2 Base model encode less cardiorespiratory information, suggesting that the shift toward linguistic representations may reduce the model's sensitivity to physiological signals. 
We also noted that larger FMs do not necessarily improve HR estimation performance, indicating that model size alone is not a reliable predictor of effectiveness in this domain. Instead, performance is significantly influenced by data splits. Figure~\ref{fig:ModelPerformance_Splits} illustrates that \texttt{Split 2} and \texttt{Split 3} result in higher errors for FM-based models, whereas the baseline method achieves better performance on these splits. Averaging predictions from FM-based models and the baseline method improves accuracy across all splits, indicating that FM representations and traditional acoustic features capture complementary cardiorespiratory information. Additionally, we observed that poor predictions often originate from the same subject recordings, with lower performance indices associated with unstable heart rates within the 5-second window, potentially due to arrhythmia or inherent annotation bias in the \emph{CirCor} Dataset, as described in Section~\ref{sec:data}.

\section{Conclusion and Future Work}\label{sec:discussion_conclusion}
In this study, we evaluated and benchmarked the ability of pre-trained acoustic foundation models (FMs) to encode heart sound auscultation for heart rate (HR) estimation. Through a layer-wise investigation of six FMs, such as HuBERT, wav2vec2, wavLM, Whisper, CLAP, and an in-house CLAP model, we demonstrated that these FMs offer comparable performance to traditional acoustic feature-based methods (using mel-spectrograms and MFCCs). We observed that the representations from the audio encoder of the in-house CLAP model achieved the lowest mean absolute error (MAE) across various data splits, outperforming the baseline model trained with standard acoustic features. Our findings indicate that acoustic foundation models, despite the domain mismatch, can be effectively adapted for auscultation and vital sign estimation, offering a robust and efficient alternative to some conventional methods. \blue{These layer-wise analysis results suggest that these FMs can be further fine-tuned to (\emph{i}) improve HR estimation by refining representation learning for cardiac acoustics and (\emph{ii}) extend their utility to clinical and pathological analysis of cardiorespiratory sounds, facilitating more accurate detection of abnormalities such as arrhythmias and murmurs.}

In the future, we plan to: (\emph{i}) explore combining acoustic features with FM representations, using feature concatenation before the downstream model or through late fusion methods within the model, for improved performance and investigate if such methods are able to capture complementary information and be more robust to individual variabilities;(\emph{ii}) investigate fine-tuning the FMs to the target domains to reduce the domain mismatch and hence explore if such adaptation translates to improved performance, \blue{better mitigate the challenges in HR estimation, and capture complex pathological characteristics}; (\emph{iii}) assess their applicability to other downstream tasks and physiological parameters, \blue{ including pathological conditions}; (\emph{iv}) augment and adapt more data \blue{that's clinically significant}; (\emph{v}) compare them with other bioacoustic foundation models, such as HeAR~\cite{baur2024hear}; and (\emph{vi}) explore model simplification strategies, such as pruning, distillation, and lightweight encoder design, to enable deployable solutions with lower computational cost while maintaining performance.







\newpage
\bibliographystyle{IEEEtran}
\bibliography{mybib}

\end{document}